\documentclass[aps,prl,showpacs,preprint]{revtex4}

\usepackage{amsmath}

\usepackage{graphicx}
\usepackage{dcolumn}
\usepackage{bm}

\begin{document}

\title{Anomalous self-diffusion in a freely evolving granular gas near the shearing instability}
\author{J. Javier Brey and M.J. Ruiz-Montero}
\affiliation{F\'{\i}sica Te\'{o}rica, Universidad de Sevilla,
Apartado de Correos 1065, E-41080, Sevilla, Spain}
\date{\today }

\begin{abstract}
The self-diffusion coefficient of a granular gas in the homogeneous cooling state is analyzed near the shearing instability. Using mode-coupling theory, it is shown that the coefficient diverges logarithmically as the instability is approached, due to the coupling of the diffusion process with the shear modes.  The divergent behavior, which is peculiar of granular gases and disappears in the elastic limit,  does not depend on any  other transport coefficient. The theoretical prediction  is confirmed by molecular dynamics simulation results for two-dimensional systems.

\end{abstract}


\maketitle

Granular gases are fluidized systems composed by particles colliding inelastically. In spite of some apparent similarities, they behave very differently from molecular fluids, and exhibit many interesting and peculiar phenomena \cite{JNyB96}. This includes the spontaneous development of strong density and temperature inhomogeneities, spontaneous symmetry breaking, pattern formation, and segregation in systems composed of different types of particles, to mention a few examples \cite{Go03}. Here, another interesting feature of granular gases is addressed, namely  the divergence of the Navier-Stokes self-diffusion coefficient when the shearing instability is approached, indicating the existence of anomalous diffusion.  This effect is quite different from the divergence of the self-diffusion coefficient in an infinite molecular two-dimensional  system, as a consequence of the algebraical long time tails of the velocity autocorrelation function (VACF) \cite{AyW67,Do75}. The divergence being discussed here occurs in rather small systems and it is due to the singular behaviour associated with the presence of a hydrodynamic instability.

In this paper, a granular gas is modeled as a system of $N$ identical smooth inelastic hard spheres or disks of mass $m$ and diameter $\sigma$, enclosed in a volume $V$, and with the collisions characterized by a constant coefficient of normal restitution $\alpha$. As a consequence of the energy dissipation in collisions, granular gases are always out of equilibrium.  For isolated systems, there is a time-dependent homogeneous cooling state (HCS), in which the granular temperature $T(t)$ decreases in time following Haff's law \cite{Ha83}, $\partial_{t}T (t)= -\zeta (t) T(t)$, $\zeta(t) \propto T(t) ^{1/2}$ being the cooling rate. This is the reference state used to derive hydrodynamic equations for granular gases \cite{BDKyS98, GyD99,BDyB08}. The predictions from these  macroscopic equations are in good agreement with simulation and experimental results for simple situations and dilute or moderately dense systems \cite{Go03}. One important result from hydrodynamics, confirmed by numerical simulations,  is that the HCS is unstable with regards to spatial perturbations with wavelength larger than some critical value, that depends on the inelasticity of the system \cite{GyZ93,McyY94}. The origin of the instability is in the fluctuations of the shear mode that lead, through nonlinear hydrodynamic contributions, to the development of density inhomogeneities \cite{BRyC99,BRyD08}.

Self-diffusion is the prototype transport process, and the associated diffusion equation is the prototype hydrodynamic equation for a macroscopic description of the process. It has been extensively investigated both in molecular and granular fluids and, in particular, in granular gases in the HCS \cite{BRCyG00,ByP00,DByL02,LByD02,ByR13}. In these studies, the hydrodynamic diffusion equation  has been derived with a prediction for the self-diffusion coefficient. It has been shown \cite{DByL02} that the coefficient, if it exists, is given by a Green-Kubo like expression. The analysis is simplified by a change from the original time scale $t$ to a new one  $s$ defined by \cite{Lu01,BRyM04}
\begin{equation}
\label{1.1}
\omega_{0} s = \ln \frac{t}{t_{0}},
\end{equation}
where $t_{0}$ and $\omega_{0}$ are two arbitrary constants. In this time scale, the original time-dependent HCS is exactly mapped on a steady state, whose granular temperature is
\begin{equation}
\label{1.2}
\widetilde{T}_{st}= \frac{m}{2} \left( \frac{\omega_{0}} {\overline{\zeta}}\right)^{2}.
\end{equation}
Here, $\overline{\zeta}$ is the time-independent  cooling rate,
\begin{equation}
\label{1.3}
\overline{\zeta} \equiv \frac{\zeta(t)}{ 2 v_{0}(t)}\, ,
\end{equation}
with $v_{0}(t) \equiv  \left( 2T/m \right)^{1/2}$. The diffusion coefficient $\widetilde{D}$ in the steady representation of the HCS is identified from the diffusion equation for the number density $n$ of tagged particles,
\begin{equation}
\label{1.4}
\frac{\partial n}{\partial s} = -  \widetilde{D}  \nabla^{2} n.
\end{equation}
At a formal level, $\widetilde{D}$ is given by the Green-Kubo expression \cite{DByL02}
\begin{equation}
\label{1.5}
\widetilde{D}= \int_{0}^{\infty} ds\, C(s),
\end{equation}
that involves the VACF in the modified dynamics  $C(s)$ for a tagged particle, defined as 
\begin{equation}
\label{1.6}
C(s) = \frac{1}{d} \langle {\bm w}(s) \cdot {\bm w}(0) \rangle,
\end{equation}
where $d$ (2 or 3) is the dimension of the system, ${\bm w}(s)$ is the velocity of the particle in the modified dynamics at time $s$, and the angular brackets denote average with the steady HCS distribution. Spatial perturbations of the transversal velocity with wavevector smaller than
\begin{equation}
\label{1.7}
k_{\perp} = \left( \frac{\widetilde{v}_{st} \overline{\zeta}}{\widetilde{\eta}} \right)^{1/2}
\end{equation}
lead to the shearing instability of the HCS mentioned above. In Eq. (\ref{1.7}), $\widetilde{v}_{st} \equiv \left( 2 \widetilde{T}_{st} /m \right)^{1/2}$ and $\widetilde{\eta}$ is related to the shear viscosity $\eta$ through
\begin{equation}
\label{1.8}
\widetilde{\eta}= \frac{\eta}{mn} \left( \frac{\widetilde{T}_{st}}{T(t)} \right)^{1/2}. 
\end{equation}
It follows that systems whose linear size $L$ is larger than $L_{c} = 2 \pi /k_{\perp}$ will exhibit the shearing instability. For this reason, the previous studies of self-diffusion in the HCS we are aware of  \cite{BRCyG00,ByP00,DByL02,LByD02,ByR13} restrict themselves to situations in which $L$ is much smaller that $L_{c}$.  On the other hand, here the focus is on the peculiarities of the process when the shearing instability is approached. The VACF can be split into a part associated to the fast relaxation of the kinetic modes, followed by the contribution of the slow relaxation of the hydrodynamic type,
\begin{equation}
\label{1.9}
C(s) = C_{\text{kin}} (s) + C_{\text{hyd}}(s).
\end{equation}
In Ref. \cite{ByR15a}, an expression for  the hydrodynamic component was derived by using mode-coupling theory. The result reads
\begin{equation}
\label{1.10}
C_{\text{hyd}}(s) \simeq \frac{(d-1) \widetilde{T}_{st}}{mnVd}  \sum_{\bm k}\nolimits  ^{(\prime)} \frac{k^{2} }{k^{2}-k_{\perp}^{2}}\, e^{-s \left[ \widetilde{\eta}\left( k^{2}-k_{\perp}^{2} \right)+k^{2} \widetilde{D} \right]} .
\end{equation}
The summation extends over  values of ${\bm k}$ compatible with the employed periodic boundary conditions, and in the interval $k_{m} \leq k \leq k_{M}$, where $k_{m}= 2 \pi/L$ and $k_{M}$ is of the order of $2 \pi$ times the inverse of the mean free path. The presence of the factor $k^{2}/(k^{2}-k_{\perp}^{2})$ is peculiar of granular gases, becoming unity in the elastic limit, in which $k_{\perp}=0$. It is associated with the existence of the shearing instability. The above expression implies that there is a time window over which the VACF exhibits a power-law decay on the $s$ scale having some similarities with the long time tails occurring in molecular fluids \cite{ByR15a}. Besides, for $ s\gg s_{0}= L^{2} /4 \pi^{2} (\widetilde{\eta} + \widetilde{D})$,  the decay of the VACF has the exponential form \cite{error}
\begin{equation}
\label{1.11}
C_{\text{hyd}} (s) \simeq \frac{2(d-1) \widetilde{T}_{st}}{mnV}  \frac{ e^{-s \left[ \widetilde{\eta}\left( k_{m}^{2}-k_{\perp}^{2} \right)+k_{m}^{2} \widetilde{D} \right]}}{1- \left( L/L_{c}\right)^{2}}\, ,
\end{equation} 
showing that the amplitude of the decay diverges as the critical length $L_{c}$ is approached. In the following, the implications of Eq. (\ref{1.10}) on the behaviour of the self-diffusion coefficient near the instability will be discussed. Let us insist on the fact that the analysis here will be for finite size $L<L_{c}$, contrary to the analysis leading to the usual long time tails in molecular gases, requiring an infinite system \cite{Do75}.  Long-time tails in freely cooling granular gases under conditions for which the HCS is unstable have been investigated in ref. \cite{HyO07}, while in ref. {\cite{FAyZ09} the case of a driven granular fluid is considered. The decomposition in Eq. (\ref{1.9}) allows us to write
\begin{equation}
\label{1.12}
\widetilde{D} = \widetilde{D}_{\text{kin}}+ \widetilde{D}_{\text{hyd}},
\end{equation}
where  $ \widetilde{D}_{\text{kin}}$ is the contribution from the kinetic part of $C(s)$ and 
\begin{equation}
\label{1.13}
\widetilde{D}_{\text{hyd}} = \frac{(d-1) \widetilde{T}_{st}}{mnVd}  \sum_{\bm k}\nolimits  ^{(\prime)} \frac{k^{2}}{(k^{2}-k_{\perp}^{2}) \left[ \widetilde{\eta} (k^{2}-k_{\perp}^{2})+k^{2} \widetilde{D} \right]}\, . 
\end{equation}
Upon deriving this expression, it has been taken into account that all the terms on the right hand side of Eq. (\ref{1.10}) decay exponentially in time if the system 
is in the parameter region in which the HCS is stable. Consider now the proximity of the shearing instability, i.e. that $L$ is close to (and below) $L_{c}$ or, equivalently, that $k_{m}$ is close to (and above) $k_{\perp}$. As long as the viscosity does not diverge as approaching the instability, it is $\widetilde{\eta} (k^{2}-k_{\perp}^{2}) \ll k^{2} \widetilde{D}$. Moreover, it will be shown below  (see Eq. (\ref{1.15})) that $\widetilde{D}_{\text{hyd}}$ diverges as the instability is approached, while 
$\widetilde{D}_{\text{kin}}$ is expected to remain finite. In this way, Eqs.\  (\ref{1.12}) and  (\ref{1.13}) in the vicinity of the instability reduce  to
\begin{equation}
\label{1.14}
\widetilde{D} \simeq \widetilde{D}_{\text{hyd}} \simeq \frac{(d-1) \widetilde{T}_{st}}{mnVd \widetilde{D}_{\text{hyd}}}  \sum_{\bm k}\nolimits  ^{(\prime)}
 \frac{1}{k^{2}-k_{\perp}^{2}}.
 \end{equation}
The summation over ${\bm k}$ can be carried out by using the continuous limit. In the following, a two-dimensional system ($d=2$) will be considered. Then, one easily gets
\begin{equation}
\label{1.15}
\widetilde{D} \simeq \widetilde{D}_{\text{hyd}} \simeq\left( \frac{\widetilde{T}_{st}}{8 \pi m n} \right)^{1/2} \left| \ln \frac{L_{c}-L}{L_{c}} \right|^{1/2} \, ,
\end{equation}
indicating a logarithmic divergent behavior of the self-diffusion coefficient as the shearing instability of the HCS is approached. Note that this behaviour differs from the one obtained by considering only the long time exponential contribution given by Eq.\ (\ref{1.11}).

Molecular dynamics (MD) simulations of a system of inelastic hard disks in a square box with periodic boundary conditions have been performed. To compare the simulation results with the above theoretical predictions, a quite precise value of the critical size  $L_{c}$ is needed for each set of parameters. In principle, $L_{c}$ is determined by Eq. (\ref{1.7}), and the expressions for the shear viscosity and the cooling rate as obtained by using the Enskog theory can be used \cite{GSyM06}. Nevertheless, those expressions  correspond to  {\em bare} quantities, i.e. neglecting hydrodynamic contributions, that can be quite relevant in the vicinity of the instability.  Then,  the critical sizes of the systems to be used in the analysis of the instability region have been identified from the simulations. A convenient method for this is based on the increase of the average kinetic energy of the system near the instability. Using fluctuating hydrodynamics, it has been shown that the average of the kinetic energy per particle $\langle e \rangle$ in the vicinity of the shearing instability behaves as \cite{BDGyM06}
\begin{equation}
\label{1.16}
\frac{\langle e \rangle -\langle e \rangle_{H}}{\langle e \rangle_{H}} = A_{e} (L-L_{c})^{-1},
\end{equation}
where $\langle e \rangle_{H}$ is the value of the average energy far away from the instability, and $A_{e}$ does not depend on $L$. This behavior is clearly observed in the MD simulations, and a fit of the data provides the value of $L_{c}$. In Fig. \ref{fig1}, the results obtained for two values of the restitution coefficient, $\alpha=0.99$ and $\alpha=0.98$, and four densities in the range $ 0.231 \leq n \sigma^{2} \leq 0.385 $ are shown. The symbols are from MD simulations and the solid lines are the theoretical predictions given by  Eq. (\ref{1.7}), using the Enskog values for the cooling rate and the shear viscosity. It is seen that the Enskog theory provides a good estimation of the critical size. A relevant conclusion of this is that the shear viscosity does not have a singular behavior when approaching the shearing instability, in agreement with the assumption made in the theory developed above.  The values of $L_{c}$ obtained in the way described will be used in the following to determine the behaviour of the self-difussion coefficient near $L=L_{c}$.
 
\begin{figure}
\includegraphics[scale=0.4,angle=0]{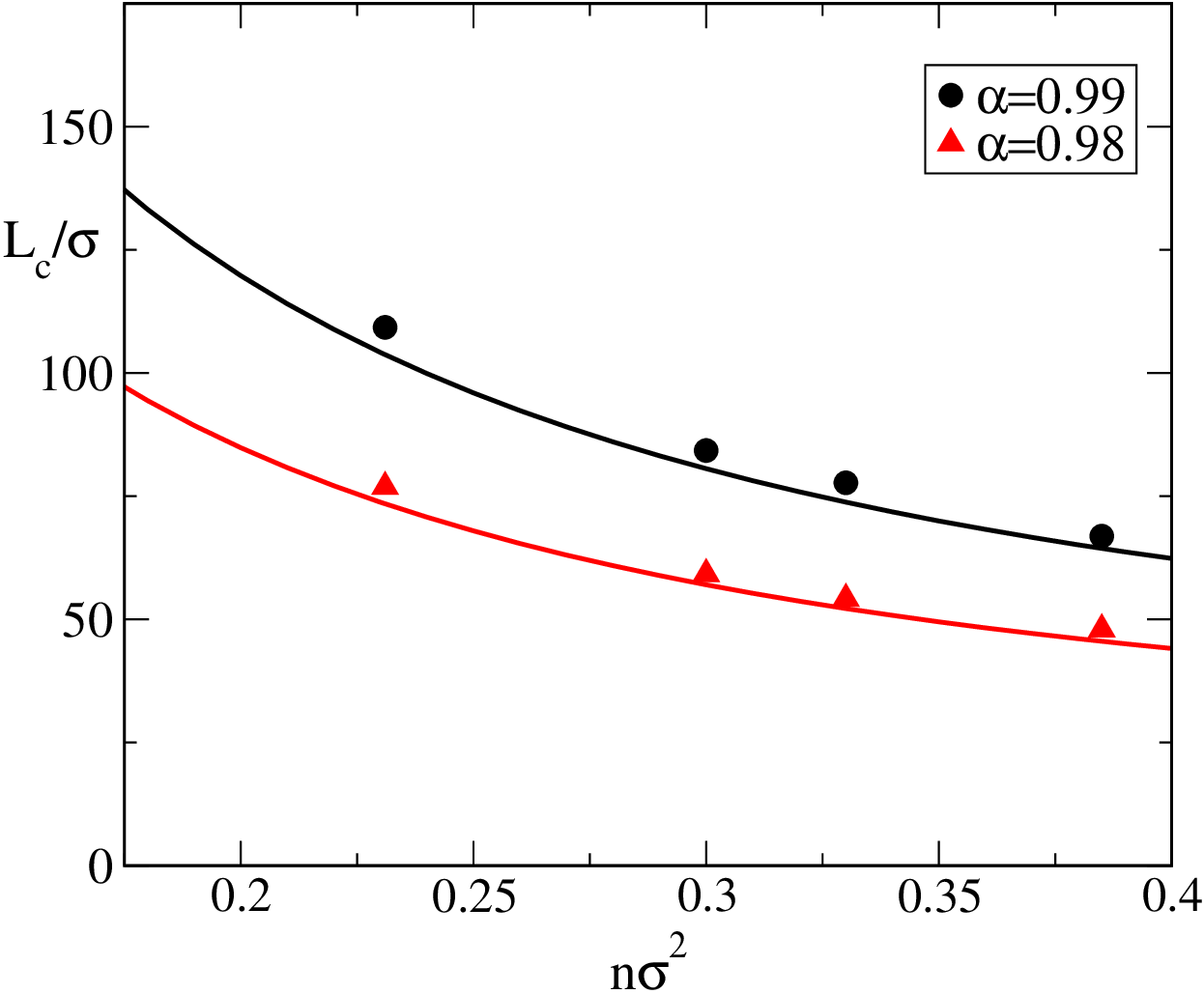}
\caption{(Color online) Critical size $L_{c}$ for the shearing instability as a function of the number density  for two values of the coefficient of normal restitution $\alpha$. The symbols are simulation results using Eq. (\ref{1.16}), while the  lines are the theoretical predictions given by Eq. (\ref{1.7}), using the Enskog values for the cooling rate and the viscosity.}
\label{fig1}
\end{figure}

The diffusion coefficient can be measured in MD simulations  by at least three different methods.  A first possibility is based on the Einstein relation for the mean square displacement of a tagged particle, that for $d=2$ is \cite{DByL02}
\begin{equation}
\label{1.17}
\widetilde{D} =\lim_{s \rightarrow \infty} \frac{1}{4s} \langle \left[ {\bm r}(s) - {\bm r}(0) \right]^{2} \rangle\, .
\end{equation}
An alternative  is to use the Green-Kubo formula, Eq.\ (\ref{1.5}).  The third way of evaluating $\widetilde{D}$ consists in considering the steady self-diffusion state reached by a mixture of two kinds of mechanically equivalent particles, differing only in some label or color, which is in contact with two reservoirs for the two types of particles. Then, the diffusion coefficient is obtained from the values of the particles flux and the concentration gradient, i.e. from  Eq.\ (\ref{1.4}) \cite{ByR13}.  The implementations of the three  methods have been discussed in detail in the literature  \cite{DByL02,ByR13}, and should lead to the same value of the coefficient of self-diffusion if the system exhibits diffusive behavior.  This consistency was confirmed in our simulations, although the results based on the VACF seem to be somewhat more accurate and less noisy, due to technical reasons, as it has been already pointed out  \cite{LByD02}. For this reason, the values being reported in the following were obtained with the Green-Kubo formula. Nevertheless, a potential difficulty arises with the tails of the VACF near the shearing instability. For large times, the decay is dominated by the hydrodynamic term with the shortest wave vector compatible with the boundary conditions, as described by Eq.\ 
(\ref{1.11}). Given that the amplitude of this contribution diverges, it could happen that the exponential long time tail would play a significant role in determining the self-diffusion coefficient. In Fig. \ref{fig2}, the long time  behavior of the VACF is illustrated for a system with $\alpha=0.98$, $n=0.3 \sigma^{-2}$, and two different system sizes. The exponential decay is clearly identified. Then what has been done, instead  of using some cut-off in the numerical integration of the VACF to get the diffusion coefficient, is to chose for each system a time $s_{E} > s_{0}$, such that the exponential decay is already clearly observed for that time. For $s<s_{E}$, the results of the simulations have been integrated numerically, while for $s_{E} < s < \infty$ what has been analytically integrated is the fit of the simulation data to an exponential, extended up to $s \rightarrow \infty$. 

\begin{figure}
\includegraphics[scale=0.4,angle=0]{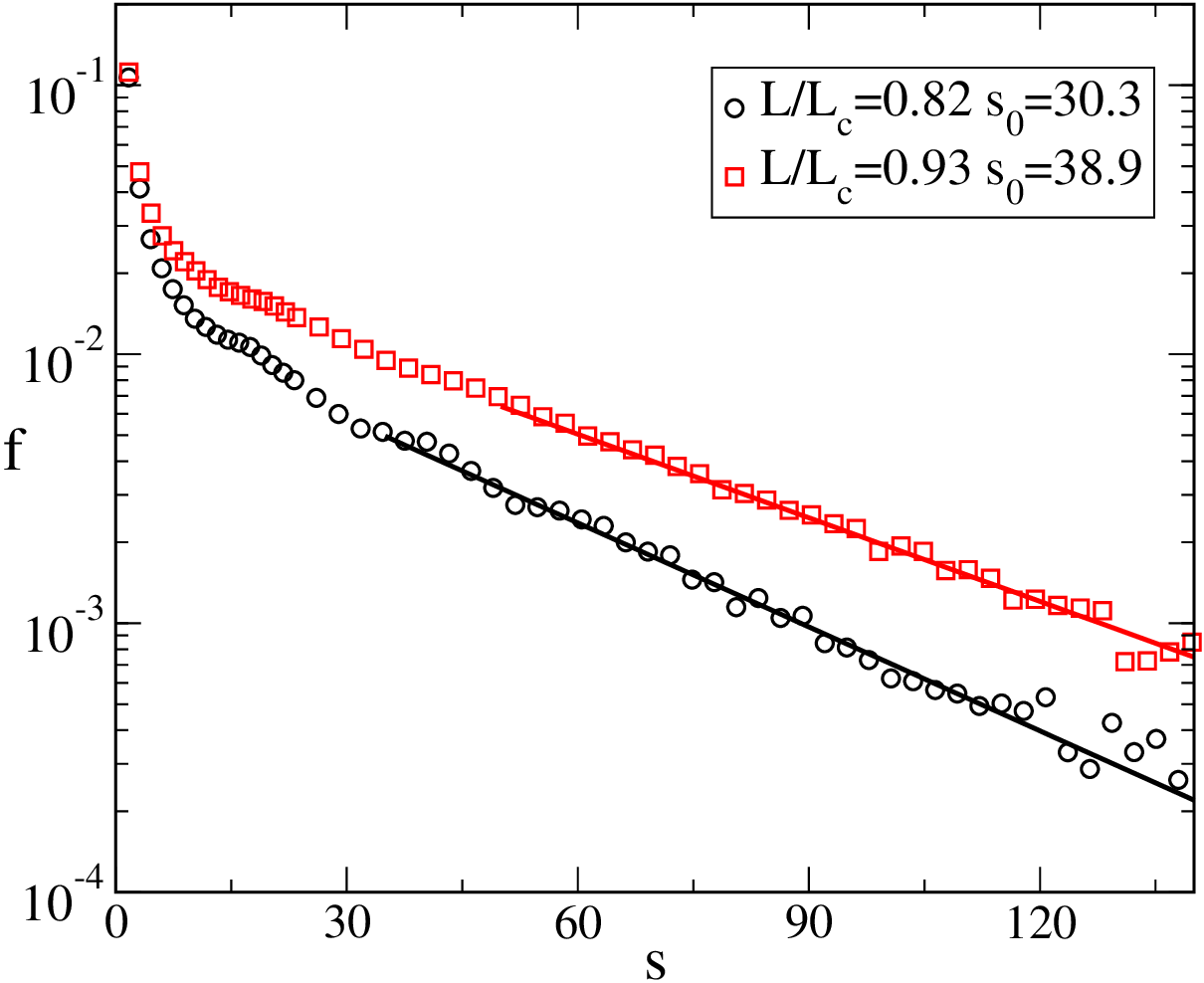}
\caption{(Color online) Time evolution of the dimensionless VACF, $f \equiv  m C(s)/ \widetilde{T}_{st}$ for a system of inelastic hard disks with $\alpha=0.98$ and $n=0.3 \sigma^{-2}$. The symbols are simulation results for two different system sizes, as indicated in the inset. The time $s_{0}$ characterizing the long time regime of Eq.\ (\ref{1.10}) is also given, The solid lines are linear fits identifying the exponential decay for long times. }
\label{fig2}
\end{figure}

In Fig. \ref{fig3} the diffusion  coefficient measured in a system with $n= 0.33 \sigma^{-2}$ and different values of the coefficient of restitution is shown as function of the size of the system. In the inelastic cases ($\alpha <1$), the divergence shows up when the size of the system approaches $L_{c}$.  Far from the divergence, $\widetilde{D}$ grows logarithmically with the size. This is the qualitative behaviour expected   in the elastic limit for all $L$, as a consequence of the long time tails \cite{AyW67}. Moreover, far from the instability the dependence of the self-diffusion coefficient on the inelasticity is rather weak. Similar results have been obtained for the other densities studied.

\begin{figure}
\includegraphics[scale=0.4,angle=0]{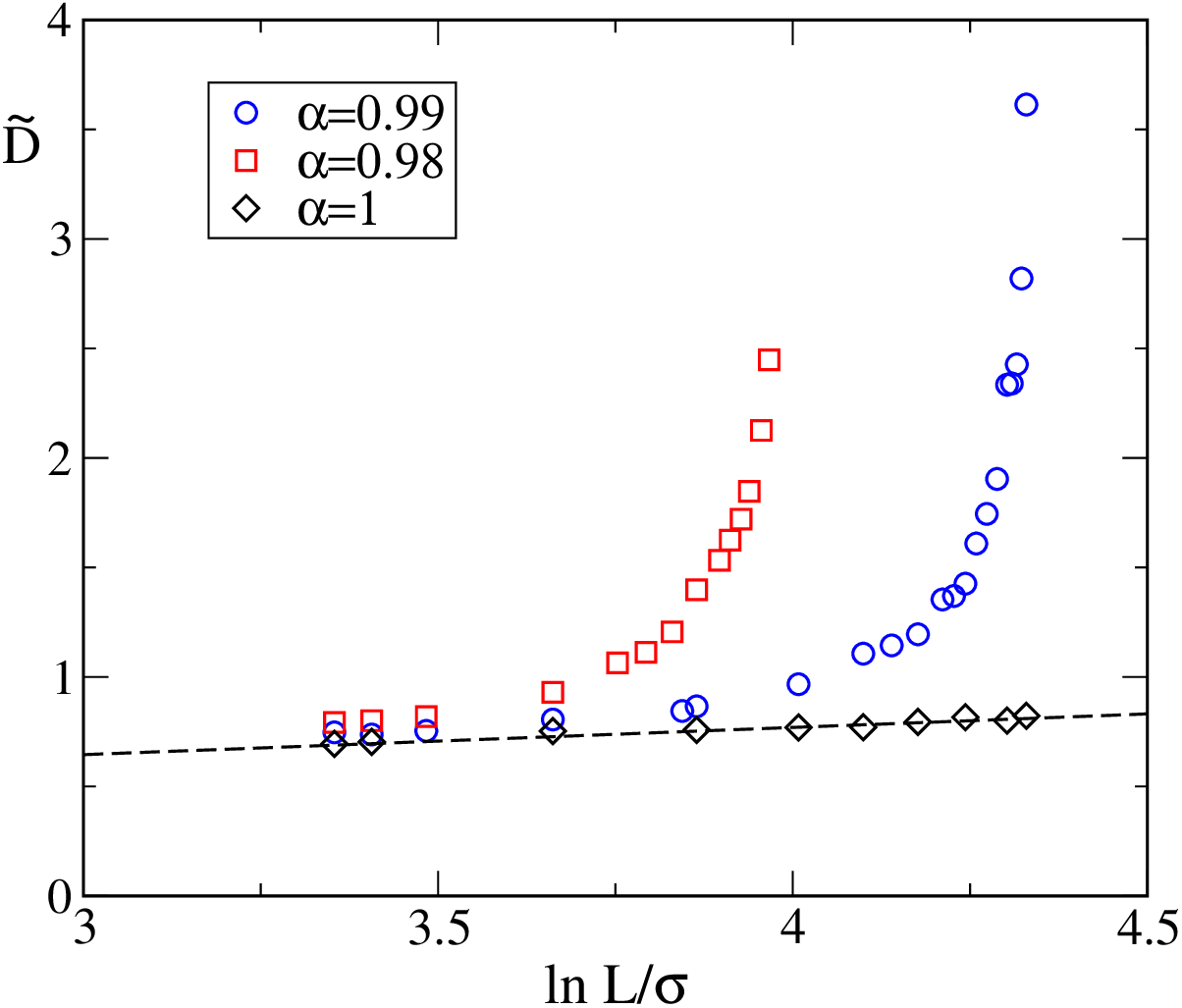}
\caption{(Color online) Dimensionless  diffusion constant $\widetilde{D}$  as a function of the size of the system $L$ for $\alpha=0.98$ (red squares), $\alpha=0.99$ (blue circles), and $\alpha =1$ (black diamonds). In all cases, it is $n=0.33 \sigma^{-2}$. The critical sizes are $\ln L_{c}/ \sigma \approx 3.99$ for $\alpha=0.98$ and $\ln L_{c}/ \sigma \approx 4.36$ for $\alpha=0.99$. The dashed line is a linear fit of the elastic data.}
\label{fig3}
\end{figure}

Also, it has been investigated whether the observed divergence is compatible with the logarithmic prediction in Eq.\ (\ref{1.15}). Then, in Fig. \ref{fig4}, $\widetilde{D}$ is plotted as a function of $[- \ln (1-L/L_{c})]^{1/2}$ for a system with $n=0.144 \sigma^{-2}$ and $\alpha=0.95$. The linear behavior predicted by the theory is  observed as the shearing instability is approached,  although on a rather limited scale range, due to the limitations imposed by the instability of the HCS itself. It is interesting to analyze the influence of the mode with the minimum wave vector $k_{m}$ on the divergence of $\widetilde{D}$.  In the same figure, the contributions to $\widetilde{D}$ from the initial decay of the VACF up to $s_{E}$ ($I_{1}$) and from the long time tail described by an exponential ($I_{2}$), as discussed above, are shown. While the former increases quite fast upon approaching $L_{c}$, the latter remains rather small. The reason is that although the amplitude of the time exponential associated to $k_{m}$ diverges, as indicated in Eq. (\ref{1.11}), its characteristic decay time goes to zero, in such a way that the time integral actually remains bounded  in the limit $L \rightarrow L_{c}$. This has been consistently observed in all the cases investigated.

In summary, using mode coupling theory the self-diffusion coefficient of a finite granular gas in the homogeneous cooling state near the shearing instability has been investigated. In the study, it turns out to be essential to use the macroscopic transport coefficients and not the {\em bare} (e.g. Enskog) ones. The result given by Eq.\ (\ref{1.14}) implies to substitute $\widetilde{D}$ by $\widetilde{D}_{\text hyd}$ on the right hand side of Eq. (\ref{1.13}), i.e. the transport coefficient considered is the (divergent) macroscopic one and not its Enskog value.  In this way, it  has been predicted that the diffusion constant exhibits a logarithmic divergence, that is consistent with the results obtained by Molecular Dynamics simulations.
The divergence is not dominated by the mode with the shortest wavevector, but by a combination of modes in the hydrodynamic region. As a by-product, it has been seen that the shear viscosity remains finite at  the shearing instability. This follows from the fact that this coeffcient appears in the hydrodynamic expression of the critical size for the system to exhibit the unstable behaviour. If the viscosity  would be infinite, the instability never would be observed in practice. The same property is crucial for the derivation of Eq. (\ref{1.15}) from Eq. (\ref{1.13}), and the latter has been shown to be consistent with the behaviour observed in MD simulations.   More extensive results will be published elsewhere.  It is worth to emphasize that the mode coupling calculation used here, relies solely on the hydrodynamic modes and, therefore, its confirmation by the simulation results is another piece of evidence for the relevance and importance of hydrodynamics in these systems, a fact often discussed in the literature.

\begin{figure}
\includegraphics[scale=0.4,angle=0]{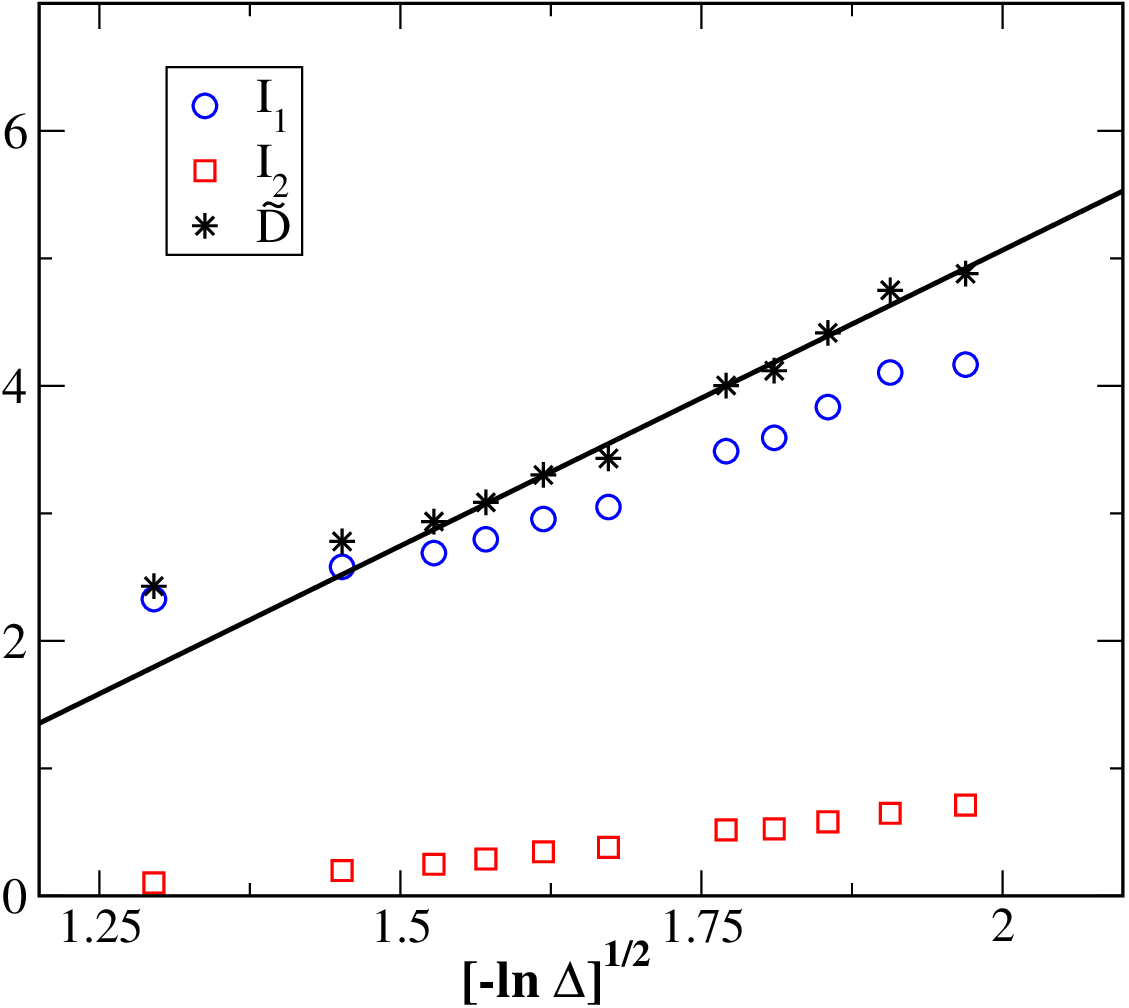}
\caption{(Color online) Dimensionless  diffusion constant $\widetilde{D}=I_{1}+I_{2}$ (black stars)  as a function of $( - \ln \Delta)^{1/2}$, where $\Delta \equiv (L_{c}-L)/L_{c}$, for a system with $\alpha=0.95$ and $n= 0.144 \sigma^{-2}$. The solid line is a linear fit corresponding to the theoretical prediction, Eq. (\ref{1.15}). The component $I_{2}$ (red squares) is the contribution from the long time exponential decay of the VACF, while $I_{1}$ (blue circles) is the contribution from shorter times. }
\label{fig4}
\end{figure}

This research was supported by the Ministerio de Econom\'{\i}a y Competitividad  (Spain) through Grant No. FIS2014-53808-P (partially financed by FEDER funds).


\begin{thebibliography}{99}
\bibitem{error} A factor of $2$ is missing in Eq. (68) of Ref. \cite{ByR15a}.
\end{thebibliography}
\end{document}